# Status of Experiment NEUTRINO-4 Search for Sterile Neutrino

A Serebrov[1], V Ivochkin[1], R Samoilov[1], A Fomin[1], A Polyushkin[1], V Zinoviev[1],
P Neustroev[1], V Golovtsov[1], A Chernyj[1], O Zherebtsov[1], V Martemyanov[2],
V Tarasenkov[2], V Aleshin[2], A Petelin[3], A Izhutov[3], A Tuzov[3], S Sazontov[3],
D Ryazanov[4], M Gromov[3], V Afanasiev[3], M Zaytsev[1, 4], M Chaikovskii[1]

[1] Petersburg Nuclear Physics Institute NRC KI, Gatchina, 188300 Russia
[2] NRC "Kurchatov institute", Moscow, 123182 Russia
[3] JSC "SSC RIAR", Dimitrovgrad, 433510 Russia
[4] DETI MEPhI, Dimitrovgrad, 433511 Russia

E-mail: serebrov@pnpi.spb.ru

**Abstract**. In order to carry out research in the field of possible existence of a sterile neutrino the laboratory based on SM-3 reactor (Dimitrovgrad, Russia) was created to search for oscillations of reactor antineutrino. The prototype of a multi-section neutrino detector with liquid scintillator volume of 350 l was installed in the middle of 2015. It is a moveable inside the passive shielding detector, which can be set at distance range from 6 to 11 meters from the reactor core. Measurements of antineutrino flux at such short distances from the reactor core are carried out with moveable detector for the first time. The measurements with full-scale detector with liquid scintillator volume of 3m$^3$ (5x10 sections) was started only in June, 2016. The today available data is presented in the article.

## 1. Introduction

At present there is a widely spread discussion about possible existence of a sterile neutrino having much less cross-section of interaction with matter, compared, for instance, with that of a reactor electron antineutrino [1, 2].

To search for neutrino oscillation in sterile state one has to observe the deviation of reactor antineutrino flux. If such a process exists it can be described by the following oscillation equation:

$$P(\tilde{\nu}_e \to \tilde{\nu}_e) = 1 - \sin^2 2\theta_{14} \ \sin^2(1.27 \frac{\Delta m_{14}^2 [eV^2] L[m]}{E_{\tilde{\nu}}[MeV]}) \qquad (1),$$

where $E_{\tilde{\nu}}$ is antineutrino energy, $\Delta m_{14}^2$ and $\sin^2 2\theta_{14}$ are the unknown oscillation parameters.

To carry out the experiment it is required to perform measurements of antineutrino flux and spectrum at short reactor distances, e.g. 6-12 meters from almost point-like antineutrino source. Due to some peculiar characteristics of its construction, reactor SM-3 provides the most favorable conditions for conducting the experiment [3]. However, SM-3 reactor is at the Earth surface, hence cosmic background is the main difficulty with this experiment. In this paper we present the results of first measurements of neutrino flux dependence on distance in baseline range 6-11 meters. All data for the prototype of a multi-section neutrino detector and today available data for the full-scale detector with liquid scintillator volume of 3m$^3$ (5x10 sections) are presented in the article.

## 2. Design of multi-section antineutrino detector model

The detector inner vessel 0.9x0.9x0.5 m$^3$ is filled with liquid scintillator doped with Gadolinium (0.1%). The scintillation type detector is based on IBD (inverse beta decay) reaction: $\tilde{\nu}_e + p \to e^+ + n$.

Detector model was designed multi-sectional to distinguish between positron emitted in inverse beta decay reaction and recoil proton from fast neutron elastic scattering. The active shielding of neutrino detector consists of external ("umbrella") and internal parts with respect to passive shielding, figure 1 (more details in [4]).

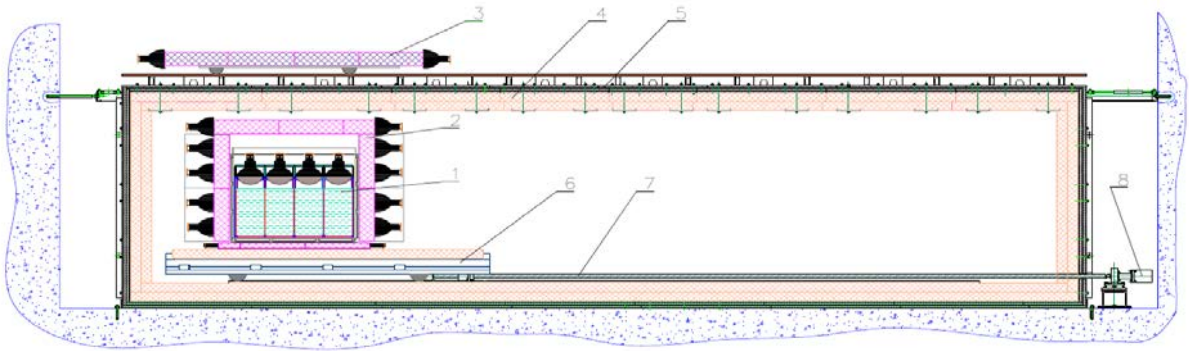

**Figure 1**. General scheme of experimental setup. 1 – detector of reactor antineutrino, 2 – internal active shielding, 3 – external active shielding (umbrella), 4 – borated polyethylene passive shielding, 5 – steel and lead passive shielding, 6 – moveable platform, 7 – feed screw, 8 – step motor.

Detector can be placed in different positions for measurement with 0.5m distance between them to avoid influence of section efficiency distinction. Shift could be made from one position to any other. The sectioned detector structure allows us to present the distance dependence with 0.5 meter step. The technique of making measurements was to move detector for 1 meter starting with the end position. On the second stage the measurements were repeated with translation of starting position for 0.5 meters. Thus, both halves of the detector measured the same point, averaging in this way somewhat different recording efficiency of each halves of the detector.

Multi-section model was designed especially for detecting positron emitted in inverse beta decay reaction. Fast neutrons from cosmic rays are the main problem for Earth-surface experiments. Fast neutron scattering imitate neutrino reaction. The recoil proton mimics the prompt signal from positron. The delayed signal emits during neutron capturing by Gd in both reactions. The difference in prompt signals is that in neutrino process two γ-quanta are emitted due to positron annihilation. The recoil proton path with high probability lies in single section. 511 keV γ-quanta can be detected in neigbouring sections. However, with section size 22.5 x 22.5 x 50 cm$^3$, about 70% prompt signals from neutrino events can be detected in single section. Hence, only ~30% of neutrino events are multi-section due to γ-quanta detection in neighbouring sections with respect to section where positron annihilated. The fact that event statistics is 3 times less if we consider only multi-section events is hardly acceptable, so we considered data analysis model with both multi and single section events. Neutrino-like events selection criteria is the ~30% to ~70% ratio for multi and single section events. Hence, if the signal difference for reactor turning on (off) is in ~30% to ~70% ratio for multi and single section events then we consider it to be neutrino-like signal.

Difference in count rates in reactor on and off regimes for double and single starts integrated over all distances was (37±4)% and (63±7)%. With considered precision this ratio allows us to regard registrated events as neutrino-like events.

## 3. Full-scale antineutrino detector

Scheme of full-scale detector is presented in figure 2. The full-scale detector with liquid scintillator has volume of 3m$^3$ (5x10 sections). The internal active shielding is located on the top of detector and under it. Due to section structure this detector allows us to carry out measurements in the similar way as for detector model but with step of 0.47m and average efficiency of different sections.

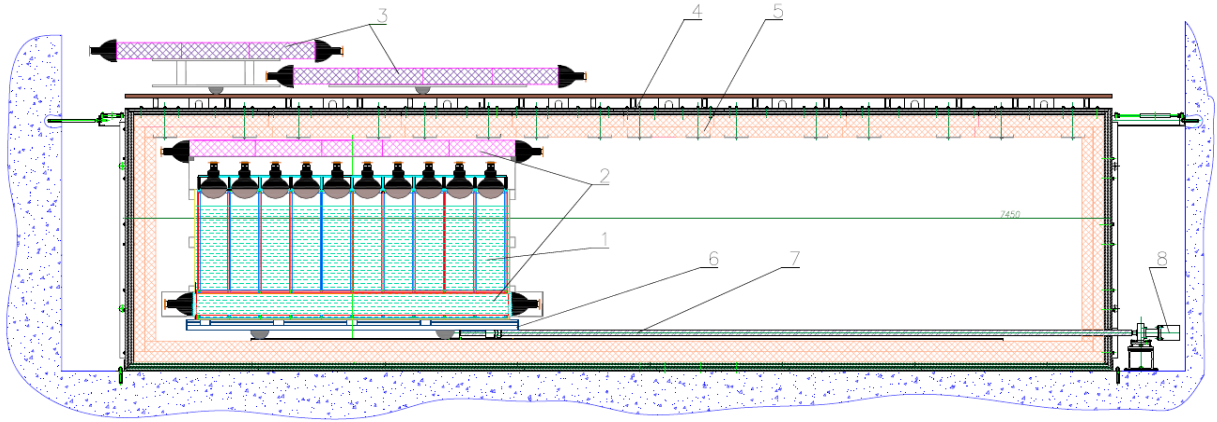

**Figure 2**. General scheme of experimental setup. 1 – detector of reactor antineutrino, 2 – internal active shielding, 3 – external active shielding (umbrella), 4 – steel and lead passive shielding, 5 – borated polyethylene passive shielding, 6 – moveable platform, 7 – feed screw, 8 – step motor.

## 4. Results

Results of measurements of difference in counting rate of neutrino-like events for model and full-scale detectors are shown in figure 3 as dependence of antineutrino on distance from reactor center.

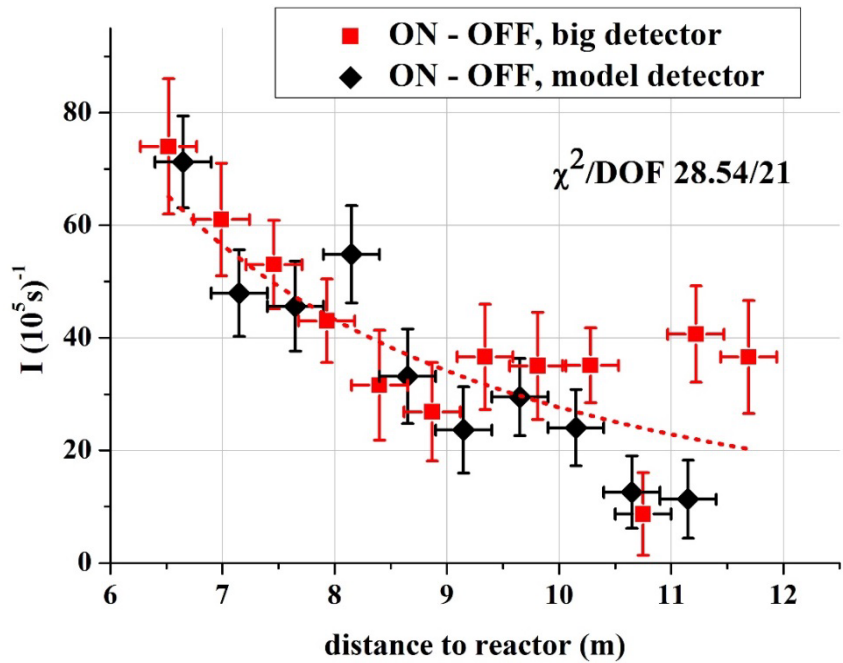

**Figure 3**. Reactor antineutrino flux distance dependence for model and full-scale detectors, point graph is the fit for dependence $1/R^2$, where R – distance from the center of reactor core.

The difference spectra (reactor ON - reactor OFF) of prompt signals with 6 distance points are presented in figure 4.

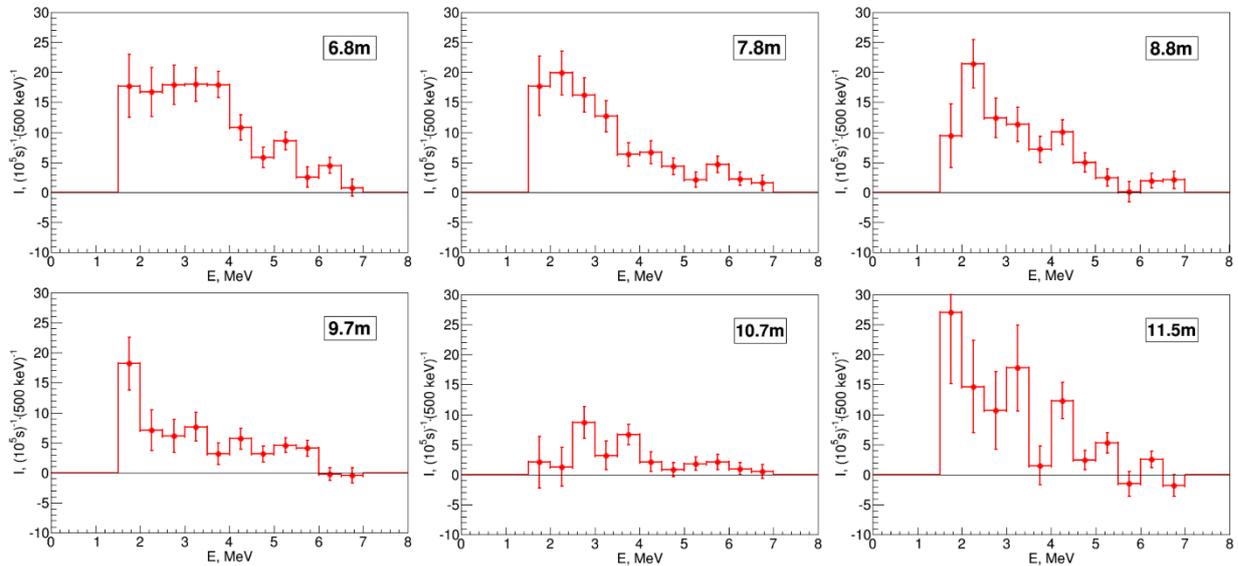

**Figure 4**. Results of spectrum measuring at various distances.

## 5. Conclusion

Measurements with full-scale detector with liquid scintillator volume of 3 m$^3$ (5x10 sections) was started only in June, 2016. First results of its work are presented. They were compared with model detector results. Measurements with new detector will be continued in the same way (movement, spectrum measurement) to reach better statistical accuracy.

Measurements of the flux of an antineutrino from the reactor at small distances of 6-12 m by means of the moveable detector are carried out for the first time. The main problem of experiment is connected with cosmic background which considerably reduces the accuracy of measurements. In the frame of the available statistical accuracy it is not revealed if there are any reliable deviations of antineutrino flux distance dependence from the law $1/R^2$ where R – distance from the center of reactor core. The results in range 10-12 m require the measurements in this region to be repeated with more accuracy.

**Acknowledgments**
The authors are grateful to the Russian Foundation for Basic Research for support under Contract No. 14-22-03055-ofi_m. The delivery of the scintillator from the laboratory leaded by Prof. Jun Cao (Institute of High Energy Physics, Beijing, China) has made a considerable contribution to this research.